\documentclass[aps,twocolumn,nofootinbib,floatfix,showpacs,superscriptaddress, 
amsmath,amssymb]{revtex4}

\usepackage{graphicx}
\usepackage{bm}
\usepackage{amssymb}
\usepackage{latexsym}
\usepackage{amsmath}

\begin{document}

%
%
%
%
\newcommand{\covdev}{{\widetilde \partial}}
\newcommand{\finalnewpage}{\newpage}
\newcommand{\lagrang}{{\cal L}}
\newcommand{\newg}{{\skew2\overline g}_\rho}
\newcommand{\Nbar}{\skew3\overline N \mkern2mu}
\newcommand{\stroke}[1]{\mbox{{$#1$}{$\!\!\!\slash \,$}}}

\newcommand{\dgamma}[1]{\gamma_{#1}}
\newcommand{\ugamma}[1]{\gamma^{#1}}
\newcommand{\dsigma}[1]{\sigma_{#1}}
\newcommand{\usigma}[1]{\sigma^{#1}}
\newcommand{\ugammafive}{\gamma^{5}}
\newcommand{\dgammafive}{\gamma_{5}}
\newcommand{\ket}[1]{\vert#1\rangle}
\newcommand{\bra}[1]{\langle#1\vert}
\newcommand{\inprod}[2]{\langle#1\vert#2\rangle}
\newcommand{\psibar}[1]{\overline{#1}}
\newcommand{\half}{\ensuremath{\frac{1}{2}}}
\newcommand{\threehalf}{\ensuremath{\frac{3}{2}}}
\newcommand{\slashed}[1]{\not\!#1}
\newcommand{\lag}{\mathcal{L}}
\newcommand{\uhalftau}[1]{\frac{\tau^{#1}}{2}}
\newcommand{\dhalftau}[1]{\frac{\tau_{#1}}{2}}
\newcommand{\ucpartial}[1]{\widetilde{\partial}^{#1}}
\newcommand{\dcpartial}[1]{\widetilde{\partial}_{#1}}
\newcommand{\mn}{\mu\nu}
\newcommand{\ab}{\alpha\beta}
\newcommand{\Tdagger}[2]{T^{\dagger \,#1}_{#2}}
\newcommand{\T}[2]{T^{#1}_{\,\,#2}}
\newcommand{\vbg}{\mathsf{v}}
\newcommand{\abg}{\mathsf{a}}
\newcommand{\sbg}{\mathsf{s}}
\newcommand{\pbg}{\mathsf{p}}
\newcommand{\modular}[2]{\vert \vec{#1}_{#2}\vert}
\def\Tr{\mathop{\rm Tr}\nolimits}

%
%
\def\dthree#1{\intback{\rm d}^3\intback #1}
\def\dthreex{\dthree{x}}
\def\fpi{f_{\pi}}
\def\gammafive{\gamma^5}
\def\gammafivel{\gamma_5}
\def\gammamu{\gamma^{\mu}}
\def\gammamul{\gamma_{\mu}}
\def\intback{\kern-.1em}
\def\kfermi{k_{\sssize {\rm F}}}    
\def\mn{{\mu\nu}}
\def\sigmamunu{\sigma^\mn}
\def\sigmamunul{\sigma_\mn}
\def\Tr{\mathop{\rm Tr}\nolimits}

\let\dsize=\displaystyle
\let\tsize=\textstyle
\let\ssize=\scriptstyle
\let\sssize\scriptscriptstyle
%
%
%


\newcommand{\beq}{\begin{equation}}
\newcommand{\eeq}{\end{equation}}
\newcommand{\beqa}{\begin{eqnarray}}
\newcommand{\eeqa}{\end{eqnarray}}

\def\Inthelimit#1{\lower1.9ex\vbox{\hbox{$\
   \buildrel{\hbox{\Large \rightarrowfill}}\over{\scriptstyle{#1}}\ $}}}
%
%

\title{Can neutrino-induced photon production \\
 explain the low energy excess in MiniBooNE?}
\author{Xilin Zhang}\email{zhangx4@ohio.edu}
\affiliation{Institute of Nuclear and Particle Physics and Department of
Physics and Astronomy, Ohio University, Athens, OH\ \ 45701, USA}

\affiliation{Department of Physics and Center for Exploration of
             Energy and Matter,\\
             Indiana University, Bloomington, IN\ \ 47405, USA}
             
\author{Brian D. Serot}\thanks{Deceased.}
\affiliation{Department of Physics and Center for Exploration of
             Energy and Matter,\\
             Indiana University, Bloomington, IN\ \ 47405, USA}

%

%
\date{\today\\[20pt]}

\begin{abstract}
This report summarizes our study of Neutral Current (NC)-induced photon production 
in MiniBooNE, as motivated by the low energy excess in this experiment [A. A. Aquilar-Arevalo \textit{et al.} (MiniBooNE Collaboration), Phys.\ Rev.\ Lett. {\bf 98}, 231801 (2007); {\bf 103}, 111801 (2009)]. It was proposed that NC photon production with two anomalous photon-$Z$ boson-vector meson couplings might explain the excess. However, our computed event numbers in both neutrino and antineutrino runs 
are consistent with the previous MiniBooNE estimate that is based on their pion 
production measurement. Various nuclear effects discussed in our previous works, 
including nucleon Fermi motion, Pauli blocking, and the $\Delta$ resonance 
broadening in the nucleus, are taken into account. Uncertainty due to the two anomalous terms and nuclear effects are studied in a conservative way.  

\end{abstract}

\smallskip

\maketitle

\section{Introduction} \label{sec:intro}
The MiniBooNE neutrino experiment searches for $\nu_{\mu}\rightarrow\nu_{e}$ and $\bar{\nu}_{\mu}\rightarrow\bar{\nu}_{e}$ oscillations \cite{MiniBNoscsum}, as motivated by the anomalous excess of $\bar{\nu}_{e}$ observed by LSND in the 
$\bar{\nu}_{\mu}$ beam \cite{LSNDoscsum}. 
Such oscillations were not clearly observed in 
MiniBooNE's neutrino run, but seem to exist in the antineutrino run, hinting at 
possible $CP$ violation.
At the same time, MiniBooNE reported a low energy excess in both runs \cite{MiniBNexcess}. It is natural to ask whether the two are interrelated. In other words, it is important to understand this excess in a complete neutrino oscillation analysis. 
For the excess, different explanations have been hypothesized, 
including anomalous photon-$Z$ boson-vector meson couplings \cite{zgammamesonints,Hillfirst,Hill2011}, Lorentz violation \cite{Kostelecky}, sterile neutrinos \cite{sterilenu}, heavy neutrino radiative decay \cite{Gninenko}, and new gauge bosons \cite{Nelson2008}. However, a clear understanding of the excess has not been achieved so far. 
It was pointed out in Ref.~\cite{MiniBNexcess} that neutral current (NC)-induced photon and $\pi^{0}$ production are potential backgrounds in the experiment, because the final photons can be misidentified as electrons in the detector. Measurement of neutrino-induced pion production has been carried out in MiniBooNE and used in their background analysis \cite{MiniBN2011pion}, but NC photon production has not been directly measured. Instead, NC photon production has been estimated by assuming a relationship between $\Delta$ radiative decay and  pion decay in the nucleus \cite{Sam12}. On the other hand, the authors in Refs.~\cite{zgammamesonints, Hillfirst, Hill2011} proposed  that anomalous photon-$Z$-meson couplings might increase NC photon production significantly to explain the excess.

A rigorous calculation of NC photon production, taking into account nuclear 
effects, has not been carried out to date but is what we aim to accomplish
in this study. There are different channels in photon production, including 
incoherent and  coherent production, but such distinction is not obvious in MiniBooNE's background  analysis. The nucleon Fermi motion, Pauli blocking, and broadening of the $\Delta$ inside nucleus [due to the opening of other decay channels (e.g. $\Delta N\rightarrow NN$)] have important effects on the production, but unfortunately are missing in the previous calculations \cite{zgammamesonints,Hillfirst, Hill2011}.  
This report summarizes our computation of NC photon event numbers in MiniBooNE. The calculation is based on a series of works dedicated to low energy (neutrino energy below $0.5$ GeV) neutrino-induced 
photon and pion production \cite{BSXZchapter, BSXZnucleon, XZBSinc, XZBScoh} 
in which the  neutrino-nucleon interaction kernel as derived from a chiral 
effective field theory (EFT) for nuclei \footnote{The EFT is knowns as quantum 
hadrodynamics EFT (QHD EFT). The motivation for this EFT and some calculated results are discussed in Refs.~\cite{SW86,SW97,Furnstahl9798,JDW04,MCINTIRE07,HU07,MCINTIRE08,BDS10}.} 
is calibrated to available pion data. Also, the nuclear effects and the 
approximation schemes in the study of incoherent and coherent production 
are benchmarked by the corresponding electron-nucleus scattering and pion photoproduction data. Two contact terms in the EFT, named as $c_{1}$ and $e_{1}$ terms here, are the low energy manifestation of the proposed anomalous photon-$Z$-meson interactions \cite{Hill2011, BSXZchapter, BSXZnucleon}. Furthermore, it was realized \cite{BSXZchapter, BSXZnucleon} that other sources also lead to these low energy interactions. In Ref.~\cite{XZBScoh}, we have calibrated the $c_{1}$ and $e_{1}$ values to the coherent pion photoproduction data, and found $c_{1}$ is correlated with the other two parameters $(r_{s}\, , \, r_{v})$ that control the $\Delta$ binding potential and spin-orbit coupling in the 
nucleus. The uncertainty of these parameters, which eventually leads to 
uncertainty in the photon event calculation, will be discussed in detail later. 

As emphasized in the previous works \cite{BSXZchapter, BSXZnucleon, XZBSinc, XZBScoh}, the EFT approach is only valid for low energy neutrinos. 
It turns out that the so-called low energy excess, if due to NC photon production, 
gets important contribution from higher energy neutrinos 
because the reconstructed neutrino energy $E_{QE}$ in MiniBooNE, based on 
quasielastic scattering kinematics, can underestimate the true energy substantially in this channel. This study essentially amounts to the high energy extrapolation of previous works \cite{BSXZchapter, BSXZnucleon, XZBSinc, XZBScoh}. To benchmark the extrapolation of the neutrino-nucleon interaction kernel, the 
neutrino-induced pion production is calculated and compared to higher energy neutrino data ($\sim 1$ GeV). A reasonable form factor is associated with $c_{1}$ and $e_{1}$ couplings to regularize their high energy behavior as inspired by the anomalous photon-$Z$-meson interactions \cite{Hill2011}. The nuclear effects are included as before. 
The final results, based on the detector information and updated detection 
efficiencies from Refs.~\cite{MiniBN2010prd, Rexdis}, indicate that in both 
neutrino and antineutrino runs, for $E_{QE}$ below $0.475$ GeV, 
the experimental estimate is close to our computation; at higher $E_{QE}$, it is close to our lower bound but significantly smaller than our upper bound,   although still compatible within the experimental uncertainty. 
We also compare our event numbers vs. $E_{\gamma}$ with the experimental estimate: our results agree with the experimental estimates, except in the lowest $E_{\gamma}$ bin ($[0.1 \ , \ 0.2]$ GeV) in both runs where ours are larger than their estimates yet the differences are not large enough to explain all the excess.     
In other words, the excess needs to be resolved from other perspectives. Furthermore, the increase 
in the number of events due to $c_{1}$ and  $e_{1}$ at higher $E_{QE}$ is more substantial in antineutrino than in neutrino data, 
which raises another interesting point: the neutrino and antineutrino respond differently to the two couplings. It is possible that, 
similar to photon production, the mechanism responsible for the excess can affect the event count differently in the two runs. Hence the statement \cite{MiniBNoscsum} should be taken cautiously that the data in the \emph{antineutrino} run above $E_{QE}=0.475$ GeV is without any unknown background solely because the \emph{neutrino} run indicates so. 

The rest of this report is organized as follows. In Sec.~\ref{sec:extrapolation}, we show our high energy extrapolation of neutrino-induced pion 
production and compare it with available data. 
After that, Sec.~\ref{sec:event} summarizes our computed photon event number 
distributions against different kinematic variables, 
and comparisons are made between them and the MiniBooNE estimates.
The uncertainty of our computation is discussed in detail there. Finally in Sec.~\ref{sec:sum}, the report ends with a short summary and discussion.

\section{High energy extrapolations and benchmarks} \label{sec:extrapolation}

\begin{figure}
\centering
\includegraphics[scale=0.5]{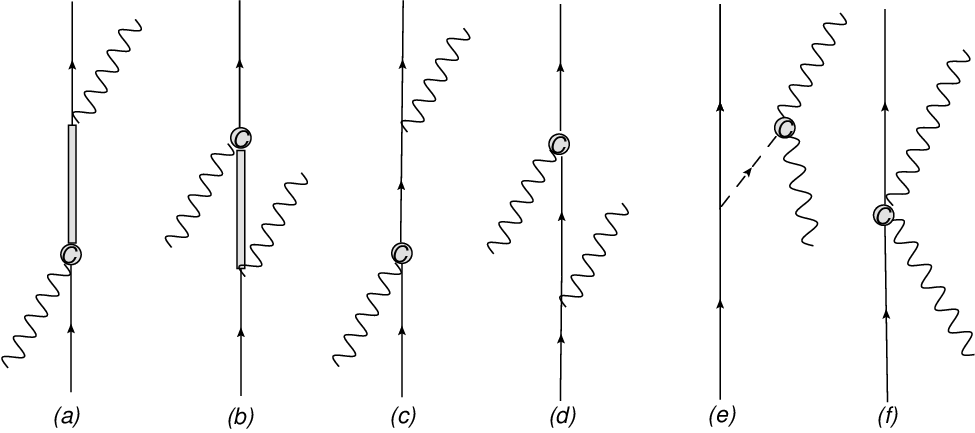}
\caption{The Feynman diagrams for photon production induced by various currents shown as ``C'' in the plot including vector, axial, and baryon currents. The diagrams for pion production can be viewed as these diagrams with final photon lines 
changed to pion lines. Diagrams (a) and (b) have the $\Delta$ as intermediate 
states; (c) and (d) have the nucleon as intermediate states; (e) has a 
pion pole in the $t$ channel; (f) includes the contact terms.}
\label{fig:feynman}
\end{figure} 

\begin{figure}
\includegraphics[scale=0.6, angle=-90]{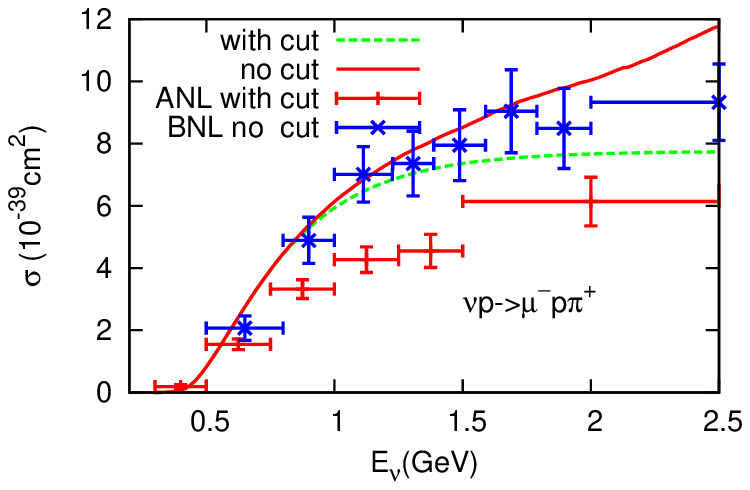}
\includegraphics[scale=0.6, angle=-90]{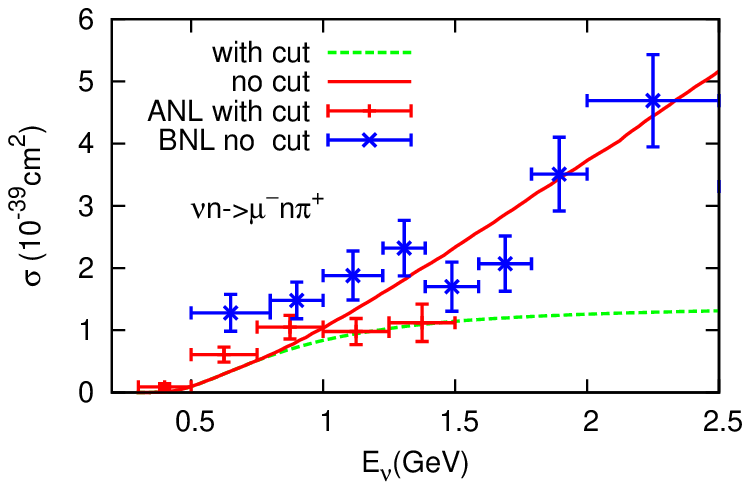}
\includegraphics[scale=0.6, angle=-90]{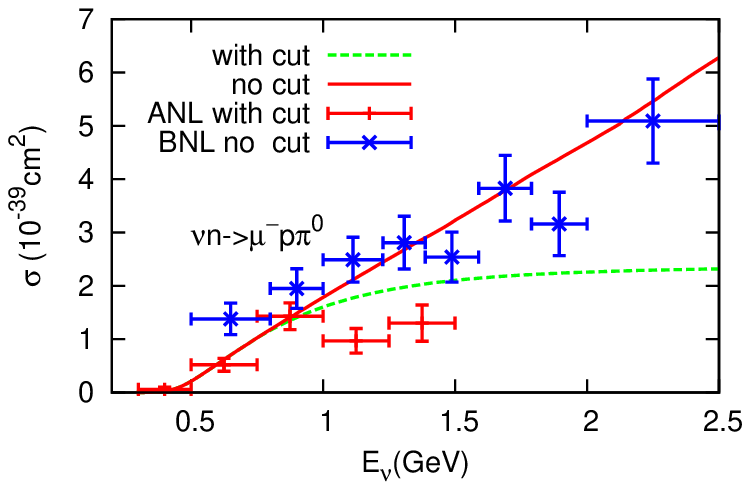}
\caption{Total cross section vs. neutrino energy in CC pion production from free nucleons. In the ANL data~\cite{RADECKY82}, the final pion and nucleon center of mass energy $M_{\pi n}$ is required to be less than $1.4 \ \mathrm{GeV}$, while no such cut is applied in the BNL data~\cite{KITAGAKI86}. Two different calculations are shown corresponding to with and without the cut.}
\label{fig:pionprodnucleon}
\end{figure}

First, to benchmark our formalism, we apply it to the study of 
neutrino-induced pion  
production from free nucleons with neutrino energy up to $\sim 2$ GeV, and check it against the existing data from ANL and BNL experiments \cite{RADECKY82,KITAGAKI86}. The contributing Feynman diagrams are similar to those in Fig.~\ref{fig:feynman} with the final photon lines changed to pion lines. 
See Refs.~\cite{BSXZchapter,BSXZnucleon} for a detailed discussion of these 
Feynman diagrams included in our previous low energy pion production study. However, only the contributions from the $\Delta$ 
resonance and nucleon intermediate states [diagrams (a)-(d)] are
considered here. Several contact terms (besides $c_{1}$ and $e_{1}$) exist, whose high energy behavior is unknown, and hence are not discussed here. To extrapolate our previous calculations \cite{BSXZnucleon, XZBSinc,XZBScoh} to higher energy, we make use of the fitted form factors of the axial transition currents ($N\leftrightarrow \Delta$) in Ref.~\cite{GRACZYK09}. 
The form factors of the vector transition currents are constrained by the  corresponding Electromagnetic interactions \cite{OL06,GRACZYK08}. The form factors of the nucleon currents [as used in diagrams (c) and (d)] are from Ref.~\cite{kelly04} (for the vector current) and Ref.~\cite{EW88} (for the axial current). Fig.~\ref{fig:pionprodnucleon} plots charged current (CC) pion production data from the ANL~\cite{RADECKY82} and BNL~\cite{KITAGAKI86} experiments. The former requires the final pion and 
nucleon center of mass energy $M_{\pi n}$ to be less than $1.4 \ \mathrm{GeV}$, but no such cut is applied in the BNL data. 
Two different calculations, with and without such a cut, are shown. We see that 
the ``no cut'' computation agrees reasonably well with the BNL data, while the ``with cut'' slightly overestimates the cross sections as indicated by the ANL data, which strongly indicates that the contact terms if regularized cannot be a dominant 
contribution in the $\sim 1$ GeV region.

\begin{figure}
\centering
\includegraphics[scale=0.6, angle=-90] {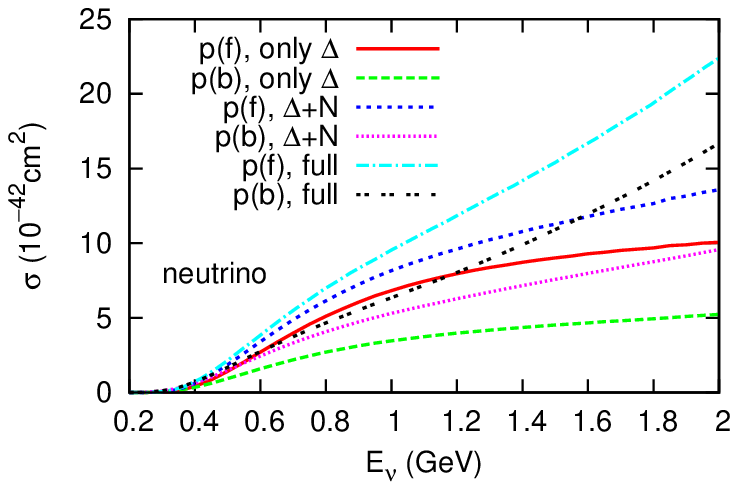}
\includegraphics[scale=0.6, angle=-90] {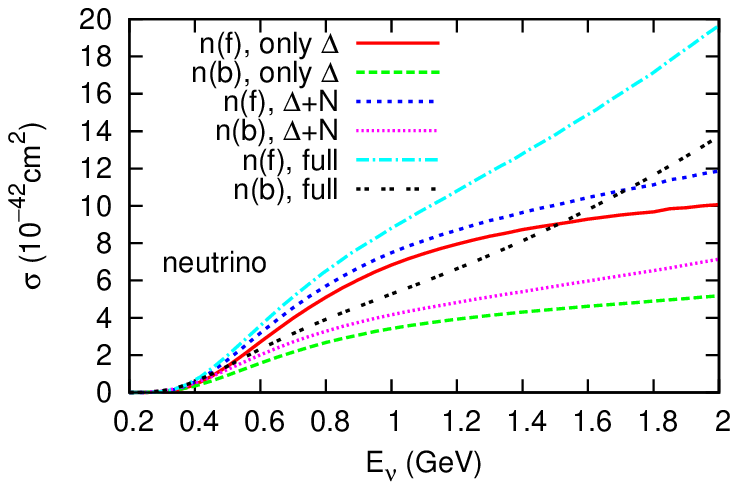}
\includegraphics[scale=0.6, angle=-90] {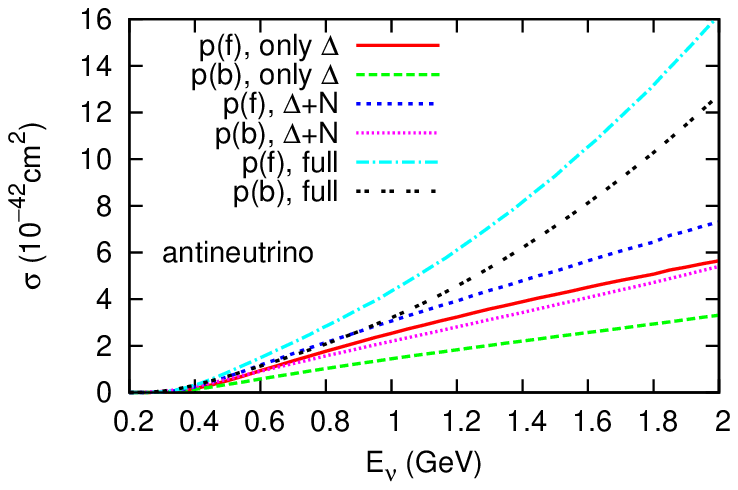}
\includegraphics[scale=0.6, angle=-90] {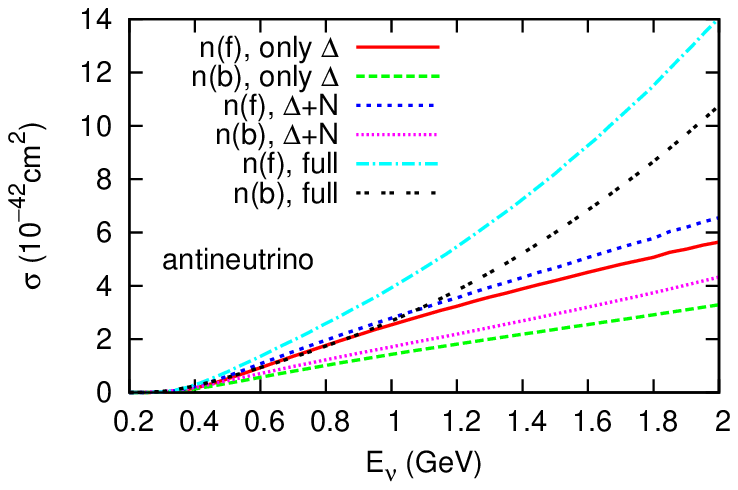}
\caption{Total cross section (per nucleon) for the photon production from free nucleons and the incoherent production from ${}^{12}C$. The first (last) two are for neutrino (antineutrino)-induced production. See the text for detailed discussion of the different curves.}
\label{fig:free_inc_xsection}
\end{figure}

Second, let's proceed to NC photon production from nucleons by using the  mentioned form factors and the procedure handling them~\cite{BSXZnucleon}. Fig.~\ref{fig:feynman} shows a list of the diagrams. However,
we can neglect the contributions from diagram (e) because of the $1-4\sin^{2}\theta_{w}$ suppression (it contributes zero in coherent production). In diagram (f), 
two couplings, the $c_{1}$ and $e_{1}$ terms, contribute here. It was pointed out 
in Ref.~\cite{Hillfirst} that the two terms can be induced by the $\omega$ and 
$\rho$ anomalous couplings with the
photon and $Z$ boson ($c_{1}\sim1.5$ and $e_{1}\sim0.8$), but also emphasized in Refs.~\cite{BSXZnucleon,Hillfirst} that other sources exist, 
including $\Delta$ off-shell interactions. At low energy, we calibrated $c_{1}$ to be around $1.5\sim 3$ by using coherent pion photoproduction data \cite{XZBScoh}. 
The two terms contribute little in the production from nucleons and the incoherent production 
with neutrino energy up to $0.5$ GeV \cite{BSXZnucleon, XZBSinc, Hillfirst}, but 
can be relevant in the coherent production \cite{XZBScoh,Hillfirst}. 
To regularize their high energy behavior, we use the following form factors as inspired by the discussion in Ref.~\cite{Hillfirst} \footnote{In the full diagram with a vector meson in the $t$ channel, there are form factors associated with the $\gamma$-$Z$-meson, and $N$-$N$-meson couplings. Between these two couplings is the meson's propagator. See Ref.~\cite{Hillfirst} for details. Here for simplicity, we parameterize these three factors in the same dipole form, and hence have the mentioned form factor in our contact diagrams.}:
\[
\left[1-\frac{q^{2}}{(1.0\ \mathrm{GeV})^{2}}\right]^{-3}  \ ,
\]
where $q$ is the change of the scattered nucleon's four momentum, i.e. $p_{f}-p_{i}$, in the interaction kernel. The total cross sections for the production from free protons and 
neutrons are shown in Fig.~\ref{fig:free_inc_xsection}. For example,  ``p(f)'' corresponds to the production from free protons, while ``p(b)'' is for the incoherent production from bound protons 
in ${}^{12}C$, which will be addressed later. In each case, three calculations including different diagrams are shown: ``only $\Delta$'' has diagram (a) in Fig.~\ref{fig:feynman}; ``$\Delta+N$'' includes (a)-(d); ``full''  includes all the diagrams [(e) is neglected as explained above]. From now on, we 
focus on the $\sim 1$ GeV region because (anti)neutrinos with higher energy are suppressed in the 
MiniBooNE spectrum. Clearly, the $\Delta$ dominates in both neutrino and antineutrino scatterings. For the proton, adding $N$ contributions increases the cross section by $\sim 10\%$, while for the neutron its contribution is less. In the ``full'' calculation, adding the contact terms, $c_{1}=3.0$ and $e_{1}=0.8$, does not change the results at low energy $\sim 0.5 $ GeV, but increases the cross section well above $\sim 1$ GeV (they are more significant in the antineutrino--induced than neutrino--induced productions). Since the contact terms and the associated 
form factors are not well understood so far, the increases well beyond $\sim 1$ GeV should be taken cautiously. 

Third, to calculate the productions in neutrino-nucleus scattering, the medium-modification of the interaction kernel needs to be included. The ground state of the nucleus (e.g. ${}^{12}C$) is computed in the mean-field approximation of QHD EFT \cite{XZBSinc, XZBScoh}. 
We apply the Local Fermi Gas approximation to calculate incoherent production 
and the ``optimal approximation'' for coherent production 
\cite{XZBSinc, XZBScoh}. The medium modifications have been checked against the inclusive electron-${}^{12}C$ scattering and coherent photoproduction data \cite{XZBSinc, XZBScoh}. In principle, the available data \cite{MiniBN2011pion} for (in)coherent 
neutrino-induced pion production from nuclei can be used to constrain theoretical models. However, before doing so, final state interaction of pions and nucleons have to be studied carefully, which is not the focus here. In addition, for 
coherent \emph{pion} production, several contact terms besides $c_{1}$ [(f) in Fig.~\ref{fig:feynman} with final photon line changed to pion line] can be important \cite{XZBScoh}, but their high energy behavior are not clear so far \cite{BSXZnucleon, HERNANDEZ07}. 
Fortunately, for photon production, the final state interaction can be ignored, and only two contact terms exists (as motivated by the anomalous couplings), which makes the high energy extrapolation more tractable for neutrino energy up to $~1$ GeV. The total cross sections for incoherent NC photon production are shown in Fig.~\ref{fig:free_inc_xsection} in comparison with those from free nucleons. 
The labeling of the different curves have already been mentioned above. 
Two parameters, $(r_{s}, \ r_{v})$, which are the meson-$\Delta$ couplings and related to the $\Delta$ binding and spin-orbit coupling, are set to be $(1, \ 1)$ in  these calculations. The incoherent production is not sensitive to different $r_{s}$ and $r_{v}$ values
around $(1, \ 1)$ (see Ref.~\cite{XZBSinc} for the details). Clearly, the $\Delta$ contribution is reduced by $50\%$ compared to that from free nucleons in all the different channels. Because of the opening of other decay channels (e.g. $\Delta N\rightarrow N N$), the $\Delta$ width increases in the nucleus. 
As a result, the flux in the radiate decay channel is decreased. Meanwhile, the nonresonant contributions from both $N$ and contact terms are reduced less, so the 
total cross sections for the full calculations are reduced less ($\sim 30\%$).  

\begin{figure}
\centering
\includegraphics[scale=0.6, angle=-90] {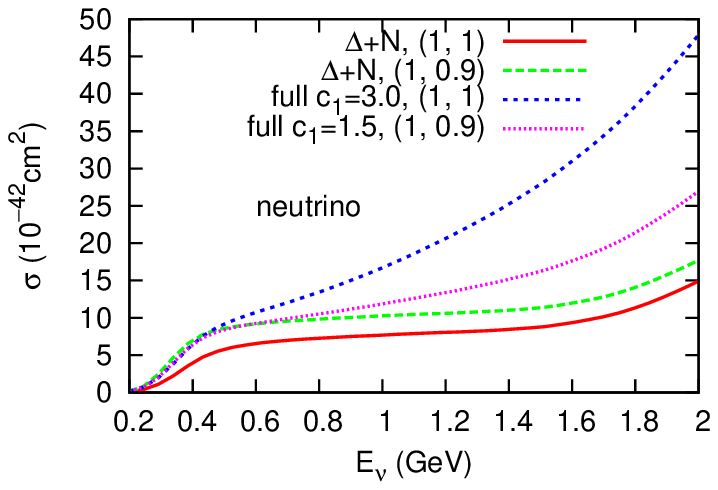}
\includegraphics[scale=0.6, angle=-90] {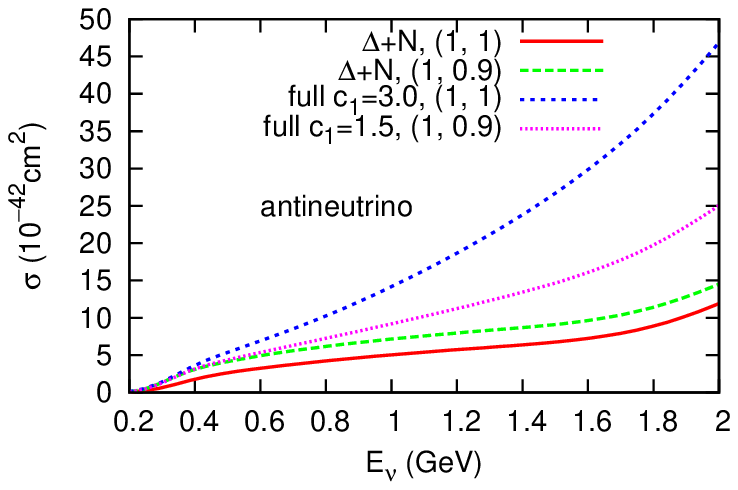}
\caption{Total cross section (per nuclei) for the coherent photon production in (anti)neutrino-${}^{12}C$ scattering. The first (last) is for neutrino (antineutrino)-induced production. See the text for detailed discussion of the
different curves.}
\label{fig:coh_xsection}
\end{figure} 

In Fig.~\ref{fig:coh_xsection}, we show the calculated total cross sections for 
coherent photon production in (anti)neutrino-${}^{12}C$ scattering. ``(1, 1)'' for example indicates $(r_{s}, \ r_{v})=(1, \ 1)$ as mentioned before.   Results of using two different sets of diagrams are compared: $\Delta+N$ includes contributions from both $\Delta$ and $N$, 
while the full calculation has all the diagrams. We can see the coherent production is indeed sensitive to $(r_{s}, \ r_{v})$ and $c_{1}$. The choices of them in the two ``full''  calculations are due to the calibration in Ref.~\cite{XZBScoh} based on coherent pion photoproduction ($e_{1}$ does not contribute in the isoscalar nucleus). 
Interestingly, for each calculation the difference between the neutrino and antineutrino-induced production cross section is reduced with increasing (anti)neutrino energy, which is consistent with previous discussion \cite{XZBScoh}. 

\section{Event numbers} \label{sec:event}

In this section, we summarize our computation of the NC photon event numbers 
in MiniBooNE. The spectrum of $\nu_{\mu}$ and $\bar{\nu}_{\mu}$ can be found in Ref.~\cite{MiniBN2010prd}: the median energy is around $0.5 \sim 1.0$ GeV, and the maximum energy is up to $\sim 2.0$ GeV. 
The proton on target (POT) numbers for the neutrino (antineutrino) run is 
$6.46 \ (11.27) \times 10^{20}$. The total effective mass of the mineral oil ($CH_{2}$) detector is $8.12 \ \times 10^{8} \ \mathrm{g}$.
The photon detection efficiency vs. photon energy is listed in 
Tab.~\ref{tab:photonefficiency}. For photon energy higher than $1.0$ GeV, we 
assume the efficiency to be $0.102$, which should not cause  mistakes because our calculations show that most photon events have energy lower than $0.8$ GeV. 

Tabs.~\ref{tab:summary_EQE} and~\ref{tab:summary_EQE_anti} list the photon event 
numbers against $E_{QE}$ (in three bins) for neutrino and antineutrino runs
(the formula for reconstructing $E_{QE}$ can be found in Ref.~\cite{MiniBN2010prd}). The events are broken into different channels including coherent (``coh'') and incoherent (``inc'') production, and the  production from unbound protons (``H''). ``$\mathrm{Total}$'' corresponds to the total number, to which MiniBooNE's estimate and excess should be compared. In each entry, we list two numbers showing the lower and upper bound of our computation. 
The lower bound in all three channels is based on the
``$\Delta+N$'' calculation with $(r_{s},\ r_{v})=(1,\ 1)$, while the upper bound 
corresponds to the 
``full calculation'' with $c_{1}=3.0, \ e_{1}=0.8,\ (r_{s},\ r_{v})=(1,\ 1)$ [for the production from free protons, $(r_{s},\ r_{v})$ are not relevant].  
See Figs.~\ref{fig:free_inc_xsection} and~\ref{fig:coh_xsection} and the text around them for a detailed discussion on the different calculations. 
Comparing different calculations in these plots, we can see that the choice of the two bounds are reasonable. In both runs for the first two $E_{QE}$ bins, 
\emph{the difference between our upper bound and experimental estimate is not significant enough to explain the excess}. In the final $E_{QE}$ bin, our upper bound is significantly larger than the estimate; 
however, this difference should be taken cautiously because our simple regularization of contact terms that are mostly responsible for the upper-lower difference have not been well benchmarked. 
Still, considering the large uncertainty of the excess in the final bin, our results are consistent with the experimental estimate indicating no excess in this bin. Another observation, as mentioned in the end of Sec.~\ref{sec:intro}, is that the contact-term induced event increase (in percentage) in the final bin is less in the neutrino run than in the antineutrino run, which means the two runs respond to the contact terms in different ways. 
 
\begin{table} 
  \centering
    \begin{tabular}{|c|c|c|c|c|c|c|c|c|} \hline
$E_{\gamma}(\mathrm{GeV}) $  & 
$0.15$  & 
$0.25$  &   
$0.35$  &      
$0.45$  &  
$0.55$  & 
$0.65$  & 
$0.75$  &  
$0.9$   \\[2pt] \hline            

 $\mathrm{Efficiency}$   & $0.089$  & $0.135$  & $0.139$  & $0.131$ & $0.123$  & $0.116$  & $0.106$  & $0.102$  \\[2pt] \hline            
     \end{tabular}
    \caption{MiniBooNE's photon detection efficiency vs. photon energy \cite{MiniBNoscsum, MiniBNdatarelase_2012}. 
The listed $E_{\gamma}$ value is the center of each energy bin. The experimental systematic uncertainty is  $15\%$.}
    \label{tab:photonefficiency}
\end{table}

\begin{table} 
    \begin{tabular}{|c|c|c|c|} \hline
$E_{QE}(\mathrm{GeV})$   & $[0.2 \, , \, 0.3]$ 
                  & $[0.3 \, , \, 0.475]$
                  & $[0.475 \, , \, 1.25]$ \\[2pt] \hline
$\mathrm{coh}$    & $1.5\ (2.9)$
                  & $6.0\ (9.2)$
                  & $2.1\ (8.0)$  \\[2pt] \hline
$\mathrm{inc}$  & $12.0\ (14.1)$
                  & $25.5\ (31.1)$
                  & $12.6 \ (23.2)$  \\[2pt] \hline  
$\mathrm{H}$  & $4.1\ (4.4)$
                  & $10.6\ (11.6)$
                  & $4.6\ (6.3)$   \\[2pt] \hline                   
$\mathrm{Total}$  & $17.6\ (21.4)$
                  & $42.1\ (51.9)$
                  & $19.3\ (37.5)$   \\[2pt] \hline
$\mathrm{MiniBN}$ & $19.5$
                  & $47.3$
                  & $19.4$   \\[2pt] \hline           
$\mathrm{Excess} $ & $42.6 \pm25.3$
                  & $82.2 \pm23.3$
                  & $21.5 \pm34.9$   \\[2pt] \hline                                                       
     \end{tabular}
    \caption{$E_{QE}$ distribution of the NC photon events in the MiniBooNE 
neutrino run, comparing our estimate to the MiniBooNE estimate \cite{MiniBNoscsum, MiniBNdatarelase_2012}.}
    \label{tab:summary_EQE}
\end{table}

\begin{table} 
    \begin{tabular}{|c|c|c|c|} \hline
$E_{QE}(\mathrm{GeV})$   & $[0.2 \, , \, 0.3]$ 
                  & $[0.3 \, , \, 0.475]$
                  & $[0.475 \, , \, 1.25]$ \\[2pt] \hline
$\mathrm{coh}$    & $1.0\ (2.2)$
                  & $3.1\ (5.5)$
                  & $0.87\ (5.4)$  \\[2pt] \hline
$\mathrm{inc}$  & $4.5\ (5.3)$
                  & $10.0\ (12.2)$
                  & $4.0\ (10.2)$  \\[2pt] \hline
$\mathrm{H}$    & $1.3\ (1.6)$
                  & $3.6\ (4.3)$
                  & $1.1\ (2.4)$   \\[2pt] \hline   
$\mathrm{Total}$  & $6.8\ (9.1)$
                  & $16.7\ (22.0)$
                  & $6.0\ (18.0)$   \\[2pt] \hline   
$\mathrm{MiniBN}$ & $8.8$
		  & $16.9$ 
		  & $6.8$     \\[2pt] \hline     
$\mathrm{Excess}$ & $34.6 \pm 13.6$
                  & $23.5 \pm 13.4$
                  & $20.2 \pm 22.8$   \\[2pt] \hline                                                      
     \end{tabular}
  
    \caption{$E_{QE}$ distribution of the NC photon events in the MiniBooNE 
antineutrino run, comparing our estimate to the MiniBooNE estimate \cite{MiniBNoscsum, MiniBNdatarelase_2012}.}
    \label{tab:summary_EQE_anti}
\end{table}

\begin{table*}
\centering
    \begin{tabular}{|c|c|c|c|c|c|c|} \hline
$E_{\gamma}(\mathrm{GeV})$  & $\mathrm{coh}$  &  $\mathrm{inc}$ 
& $\mathrm{H}$  & $\mathrm{Total}$  & $\mathrm{MiniBN}$ &  $\mathrm{Excess} $ \\[2pt] \hline                                      
 $[0.1 \, , \, 0.2]$ &$0.72\ (1.5)$ 
                     &$14.0 \ (15.0)$ 
                     & $4.4\ (4.6)$ 
                     & $19.1\ (21.1)$
                     & $10.6$
                     & $52.5$   \\[2pt] \hline 
 $[0.2 \, , \, 0.3]$ & $3.2\ (5.5)$ 
                     & $22.7 \ (25.2)$
                     & $7.8\ (8.5)$ 
                     & $33.7 \ (39.2)$
                     & $32.5$
                     & $61.2$   \\[2pt] \hline 
 $[0.3 \, , \, 0.4]$ & $3.7\ (5.4)$ 
                     & $12.7 \ (15.0)$
                     & $5.0\ (5.6)$ 
                     & $21.4 \ (26.0)$
                     & $24.7$
                     & $58.4$   \\[2pt] \hline 
 $[0.4 \, , \, 0.5]$ & $1.0\ (1.7)$ 
                     & $5.4\ (7.3)$
                     & $2.1\ (2.4)$ 
                     & $8.5 \ (11.4)$
                     & $12.7$
                     & $-9.7$   \\[2pt] \hline 
 $[0.5 \, , \, 0.6]$ & $0.32\ (1.0)$ 
                     & $2.3 \ (3.9)$
                     & $0.75\ (1.0)$
                     & $3.4\ (5.9)$
                     & $4.4$
                     & $10.5$   \\[2pt] \hline        
     \end{tabular}
    \caption{$E_{\gamma}$ distribution of the NC photon events in the MiniBooNE 
neutrino run, comparing our estimate to the MiniBooNE estimate \cite{MiniBNoscsum, MiniBNdatarelase_2012}.}
    \label{tab:summary_Egamma}
\end{table*}

\begin{table*} 
\centering
    \begin{tabular}{|c|c|c|c|c|c|c|} \hline
$E_{\gamma}(\mathrm{GeV})$  & $\mathrm{coh}$  &  $\mathrm{inc}$ 
& $\mathrm{H}$  & $\mathrm{Total}$  & $\mathrm{MiniBN}$ &  $\mathrm{Excess} $ \\[2pt] \hline                                      
 $[0.1 \, , \, 0.2]$ & $0.55\ (1.2)$ 
                     & $4.9 \ (5.5)$ 
                     & $1.4\ (1.6)$
                     & $6.9\ (8.3)$
                     & $4.3$
                     & $18.8$   \\[2pt] \hline 
 $[0.2 \, , \, 0.3]$ & $2.0\ (3.8) $ 
                     & $8.7 \ (10.3)$ 
                     & $2.9\ (3.3)$
                     & $13.6\ (17.4)$
                     & $14.3$
                     & $22.6$   \\[2pt] \hline 
 $[0.3 \, , \, 0.4]$ & $1.8\ (3.0) $ 
                     & $4.0 \ (5.4)$
                     & $1.5\ (1.8)$
                     & $7.3 \ (10.2)$
                     & $9.1$
                     & $11.5$   \\[2pt] \hline 
 $[0.4 \, , \, 0.5]$ & $0.36\ (1.0)$ 
                     & $1.3 \ (2.6)$
                     & $0.43\ (0.66)$
                     & $2.1\ (4.3)$
                     & $3.6$
                     & $18.7$   \\[2pt] \hline 
 $[0.5 \, , \, 0.6]$ & $0.10\ (0.72)$ 
                     & $0.51 \ (1.7)$
                     & $0.14\ (0.36)$
                     & $0.75 \ (2.8)$
                     & $1.1$
                     & $8.4$   \\[2pt] \hline        
     \end{tabular}
   \caption{$E_{\gamma}$ distribution of the NC photon events in the MiniBooNE 
antineutrino run, comparing our estimate to the MiniBooNE estimate \cite{MiniBNoscsum, MiniBNdatarelase_2012}.  }
    \label{tab:summary_Egamma_anti}
\end{table*}

Tabs.~\ref{tab:summary_Egamma} and~\ref{tab:summary_Egamma_anti} show the same 
event numbers as in Tabs.~\ref{tab:summary_EQE} and~\ref{tab:summary_EQE_anti} 
but against a different variable, photon energy $E_{\gamma}$, 
in five different bins. The labeling has been mentioned above. We can see that the experimental estimates agree reasonably well with our results, except for the lowest $E_{\gamma}$ bin ($[0.1 \ , \ 0.2]$ GeV) in which their estimates fall below our lower bounds in both neutrino and antineutrino runs. This is perhaps due to our inclusion of nonresonant diagram contributions, which is not manifestly included in the experimental estimates.  
 
Comparing our results with those in Ref.~\cite{Hill2011}, we can see that 
with the updated photon detection efficiencies used here, 
our results are 
generally smaller than those in Ref.~\cite{Hill2011}. In Sec.~\ref{sec:extrapolation}, we have shown that nuclear effects tend to reduce the total cross section 
substantially, which is also reflected in the event number computation.

\section{summary} \label{sec:sum}
In summary, we have calculated 
the total cross sections for NC photon production and 
computed the photon production event numbers vs. reconstructed neutrino energy 
$E_{QE}$ and photon energy $E_{\gamma}$ for MiniBooNE. 
Two contact terms $c_{1}$ and $e_{1}$ (with reasonable form factors) that are partially related to the proposed anomalous photon-$Z$-meson couplings are also discussed here. 
The other two parameters $(r_{s},\ r_{v})$ that control the  $\Delta$ binding and spin-orbit coupling in the nucleus are another source of uncertainty in our 
calculations. Previously based on inclusive electron-nucleus scattering data and 
phenomenological fits of the $\Delta$ spin-orbit coupling in the study of 
pion-nucleus scattering, we set $(r_{s},\ r_{v})$ to be around $(1, \ 1)$. The further study of coherent pion photoproduction shows fixing $(r_{s},\ r_{v})$ is entangled with $c_{1}$: Two sets, $(r_{s},\ r_{v})=(1, \ 1), \ c_{1}=3.0$ and $(r_{s},\ r_{v})=(1, \ 0.9), \ c_{1}=1.5$, are allowed. 

To test the neutrino-nucleon interaction kernel with neutrino energy $\sim 1$ GeV, the corresponding pion production with $\Delta$ and $N$ contributions included has been shown to be in good agreement with available data. Based on the transition and other form factors as benchmarked in pion production, we calculate the total cross 
section for photon production, including contributions from free nucleons, 
incoherent production, and coherent production in the high neutrino energy region. As we can see, the regularized $c_{1}$ and $e_{1}$ contributions are relatively small compared to $\Delta+N$ in all the photon production channels until neutrino energy goes beyond $\sim 1 $ GeV, which indicates our full calculation with the two contact terms provides a conservative upper bound in the event count. We treat the $\Delta+N$ contribution as the lower bound. Based on the two bounds, we conclude that the difference (our count $-$ MiniBooNE's estimate) is not significant enough to explain the excess with  $E_{QE}$ below $0.475$ GeV in both neutrino and antineutrino runs. For  $E_{QE}$ above $0.475$ GeV, although our result (upper bound) is significantly larger than the estimate, it is still compatible with no excess within the given uncertainty range. For the $E_{\gamma}$ distribution, again the experimental estimates agree with our results, except that in the lowest bins in both runs, ours are larger than their estimates but the differences could not explain all the excess in those bins.
Hence, other perspectives need to be explored to understand the low energy excess 
in MiniBooNE.  
In addition, the two contact terms provide an exemplary mechanism to which the neutrino and antineutrino runs in MiniBooNE respond differently. 
As a result, the nonexistence of low energy excess beyond $0.475$ GeV in the former does not immediately indicate it is also true in the latter and vice versa. 

To eliminate the uncertainty in the neutrino-nucleon interaction kernel, a model independent approach needs to be implemented, for example the dispersion analysis, which is currently begin pursued. Moreover, this is also relevant to the known $\gamma $ $Z$ box diagram in the study of parity violation of electron-nucleon (nucleus) scattering. 

\section*{Acknowledgements}
X.Z. would like to thank Charles J. Horowitz for introducing him the low energy excess at MiniBooNE, and thank Gerald T. Garvey, Joe Grange,  Teppei Katori, William C. Louis, Rex Tayloe, and Geralyn Zeller for their valuable information about MiniBooNE and useful discussions. This work was mainly supported by the Department of Energy under Contract No.\ DE--FG02--87ER40365. The supports from the Nuclear Theory Center at Indiana University and the US Department of Energy under grant DE-FG02-93ER-40756 are also acknowledged.


\begin{thebibliography}{100}

\bibitem{MiniBNoscsum}
A.~A.~Aguilar-Arevalo \emph{et al.}, (MiniBooNE Collaboration), Phys.\ Rev.\ 
Lett. {\bf 98}, 231801 (2007); {\bf 103}, 111801 (2009); arXiv:1207.4809v2 [hep-ex]. 

\bibitem{LSNDoscsum}
C.~Athanassopoulos \emph{et al.}, Phys.\ Rev.\ Lett. {\bf 75}, 2650
(1995); {\bf 77}, 3082 (1996); {\bf 81, 1774} (1998); A.\ Aguilar \emph{et al.}, Phys.\ Rev.\ D {\bf 64}, 112007 (2001).

\bibitem{MiniBNexcess}
A.~A.~Aguilar-Arevalo \emph{et al.}, (MiniBooNE Collaboration), Phys.\ Rev.\ 
Lett. {\bf 102}, 101802 (2009);  {\bf 105}, 181801 (2010).

\bibitem{zgammamesonints}
S.~S.~Gershtein, Yu.~Ya.~Komachenko, and M.\ Yu.\ Khlopov, 
Sov.\ J.\ Nucl.\ Phys.\ {\bf 33}, (1981) 860; J.~A.~Harvey, C.~T.~Hill, and R.~J.~Hill, Phys.\ Rev.\ Lett. {\bf 99}, 261601 (2007); Phys.\ Rev.\ D {\bf 77}, 085017 (2008).

\bibitem{Hillfirst}
R.~J.~Hill, Phys.\ Rev.\ D {\bf 81}, 013008 (2010).

\bibitem{Hill2011}
R.~J.~Hill, Phys.\ Rev.\ D {\bf 84}, 017501, (2011).

\bibitem{Kostelecky}
V.~A.~Kostelecky and M.~Mewes, Phys.\ Rev.\ D {\bf 69}, 016005
(2004); T.~Katori, V.~A.~Kostelecky, and R.~Tayloe, Phys.\ 
Rev.\ D {\bf 74}, 105009 (2006).

\bibitem{sterilenu}
M.~Sorel, J.~M.~Conrad, and M.~H.~Shaevitz, Phys.\ Rev.\ D {\bf 70}, 073004 (2004); G.~Karagiorgi \emph{et al.}, Phys.\ Rev.\ D {\bf 75}, 013011 (2007); Alessandro Melchiorri \emph{et al.}, JCAP, {\bf 0901}, 36 (2009); 
Heinrich Pas, Sandip Pakvasa, and Thomas J.~Weiler, Phys.\ Rev.\ D {\bf 72}, 095017 (2005); T.~Goldman, G.~J.~Stephenson, Jr., and B.~H.~J.~McKellar, Phys.\ Rev.\ D {\bf 75}, 091301(R) (2007); Michele Maltoni and Thomas Schwetz, Phys.\ Rev.\ D {\bf 76}, 093005 (2007).

\bibitem{Gninenko}
S.~N.~Gninenko, Phys.\ Rev.\ Lett. {\bf 103}, 241802 (2009); Phys.\ Rev.\ D {\bf 83}, 015015 (2011).

\bibitem{Nelson2008}
Ann~E.~Nelson and Jonathan Walsh, Phys.\ Rev.\ D {\bf 77}, 033001 (2008).

\bibitem{MiniBN2011pion}
A. A. Aguilar-Arevalo \emph{et al.}, (MiniBooNE Collaboration), Phys.\ Rev.\ D {\bf 81}, 013005 (2010); {\bf 83}, 052007 (2011); {\bf 83}, 052009 (2011); Phys.\ Lett.\ B {\bf 664}, 41 (2008).

\bibitem{Sam12}
Geralyn Zeller and G.~T.~Garvey, (private communications).

\bibitem{BSXZchapter}
B. D. Serot and X. Zhang, arXiv:1011.5913v1 (nucl-th); \emph{Advances in Quantum Field Theory}, Sergey Ketov, ed. (InTech, Croatia, 2012), Ch. 4 [arXiv:1110.2760v1 (nucl-th)].  

\bibitem{BSXZnucleon} B. D. Serot and X. Zhang, Phys.\ Rev.\ C {\bf 86}, 015501 (2012) [arXiv:1206.3812v1 (nucl-th)].

\bibitem{XZBSinc} X. Zhang and B. D. Serot, Phys.\ Rev.\ C  {\bf 86}, 035502 (2012) [arXiv:1206.6324v1 (nucl-th)].

\bibitem{XZBScoh} X. Zhang and B. D. Serot, Phys.\ Rev.\ C {\bf 86}, 035504 (2012) [arXiv:1208.1553v1 (nucl-th)].

\bibitem{SW86} B.~D. Serot and J.~D. Walecka, Adv.\ Nucl.\ Phys.~{\bf
               16}, 1 (1986).
%
\bibitem{SW97}B.~D.~Serot and J.~D.~Walecka,
  Int.\ J.\ Mod.\ Phys.\ E {\bf 6}, 515 (1997).
%
\bibitem{Furnstahl9798}R.~J. Furnstahl, B.~D.~Serot, and H.-B.~Tang,
  Nucl.\ Phys.\ A {\bf 615}, 441 (1997); {\bf 640}, 505 (E) (1998).
  
\bibitem{MCINTIRE07} J. McIntire, Y. Hu, and B. D. Serot, Nucl.\
              Phys.\ A {\bf 794}, 166 (2007).
%
\bibitem{HU07} Y. Hu, J. McIntire, and B. D. Serot, Nucl.\ Phys.\ A
               {\bf 794}, 187 (2007). 
%
\bibitem{MCINTIRE08} J. McIntire, Ann.\ of Phys.\ {\bf 323}, 1460
                   (2008).
%
\bibitem{JDW04} J. D. Walecka, Theoretical Nuclear and Subnuclear
              Physics, second ed. (World Scientific, Singapore,
              2004), Ch.~24.
%
\bibitem{BDS10} B. D. Serot, Phys.\ Rev.\ C {\bf 81}, 034305 (2010).

\bibitem{MiniBN2010prd}
A. A. Aguilar-Arevalo \emph{et al.}, (MiniBooNE Collaboration) Phys.\ Rev.\ 
D. {\bf 81}, 092005 (2010)
 

\bibitem{Rexdis}
R.~Tayloe, (private communication). 

\bibitem{RADECKY82} G. M. Radecky \emph{et al.}, Phys.\ Rev.\ D {\bf
                 25}, 1161 (1982).
%
\bibitem{KITAGAKI86} T. Kitagaki \emph{et al.}, Phys.\ Rev.\ D {\bf 34}, 2554 (1986).

\bibitem{OL06} O. Lalakulich, E. A. Paschos, and G. Piranishvili,
               Phys.\ Rev.\ D {\bf 74}, 014009 (2006).
       
\bibitem{GRACZYK08} K. M. Graczyk and J. T. Sobczyk, Phys.\ Rev.\ D
               {\bf 77}, 053001 (2008).       

\bibitem{GRACZYK09} K. M. Graczyk, D. Kie{\l}czewska, P.
                 Przew{\l}ocki, and J. T. Sobczyk, Phys.\ Rev.\ D
                 {\bf 80}, 093001 (2009).

\bibitem{kelly04} J. J. Kelly, Phys.\ Rev.\ C {\bf 70}, (2004) 068202.
%
\bibitem{EW88} T. Ericson and W. Weise, Pions and Nuclei (Clarendon
               Press, Oxford, 1988).

\bibitem{HERNANDEZ07} E. Hern{\'a}ndez, J. Nieves, and M. Valverde,
               Phys.\ Rev.\ D {\bf 76}, 033005 (2007).  
%
\bibitem{MiniBNdatarelase_2012} \url{http://www-boone.fnal.gov/for_physicists/data_release/nue_nuebar_2012/} .               
\end{thebibliography}
\end{document}